\documentclass{aa} 
\usepackage{txfonts,epsfig,graphicx,natbib,url,twoopt}

\newcommand{\mubold}{\mbox{\boldmath$\mu$}}
\newcommand{\thetabold}{\mbox{\boldmath$\theta$}}

\newcommand{\phibold}{\mbox{\boldmath$\phi$}}
\newcommand{\epsilonbold}{\mbox{\boldmath$\epsilon$}}

\usepackage[breaklinks=true]{hyperref} 
\bibpunct{(}{)}{;}{a}{}{,}             
\makeatletter
  \newcommandtwoopt{\citeads}[3][][]{\href{http://adsabs.harvard.edu/abs/#3}%
    {\def\hyper@linkstart##1##2{}%
     \let\hyper@linkend\@empty\citealp[#1][#2]{#3}}}
  \newcommandtwoopt{\citepads}[3][][]{\href{http://adsabs.harvard.edu/abs/#3}%
    {\def\hyper@linkstart##1##2{}%
     \let\hyper@linkend\@empty\citep[#1][#2]{#3}}}
  \newcommandtwoopt{\citetads}[3][][]{\href{http://adsabs.harvard.edu/abs/#3}%
    {\def\hyper@linkstart##1##2{}%
     \let\hyper@linkend\@empty\citet[#1][#2]{#3}}}
  \newcommandtwoopt{\citeyearads}[3][][]%
    {\href{http://adsabs.harvard.edu/abs/#3}
    {\def\hyper@linkstart##1##2{}%
     \let\hyper@linkend\@empty\citeyear[#1][#2]{#3}}}
\makeatother

\begin{document}

   \title{Hierarchical analysis of the quiet Sun magnetism}

   \author{A. Asensio Ramos \and M. J. Mart\'{\i}nez Gonz\'alez
          }

   \institute{Instituto de Astrof\'\i sica de Canarias,
              38205, La Laguna, Tenerife, Spain; \email{aasensio@iac.es}
            \and
Departamento de Astrof\'{\i}sica, Universidad de La Laguna, E-38205 La Laguna, Tenerife, Spain
             }

  \abstract{Standard statistical analysis of the magnetic properties of the quiet Sun rely on simple histograms of 
  quantities inferred from maximum-likelihood estimations. Because of the inherent degeneracies, either intrinsic or induced by the
  noise, this approach is not optimal and can lead to highly biased results. We carry out a meta-analysis of the 
  magnetism of the quiet Sun from Hinode observations using a hierarchical probabilistic method. This model allows us to infer the
  statistical properties of the magnetic field vector over the observed field-of-view consistently taking into account the uncertainties in
  each pixel due to noise and degeneracies. Our results point out
  that the magnetic fields are very weak, below 275 G with 95\% credibility, with a slight preference for horizontal fields, although the distribution is
  not far from a quasi-isotropic distribution.}

   \authorrunning{Asensio Ramos \& Mart\'{\i}nez Gonz\'alez}
   \maketitle
%

\section{Introduction}

One of the most interesting advances on the study of the magnetism of the Sun is 
the relatively recent
observation of a small-scale, very dynamic magnetism that pervades the quietest 
areas of the solar surface. This magnetism was first characterized through the 
investigation of
the polarimetric signals produced by the Zeeman effect in the near-infrared 
\citep{lin95, khomenko03} or simultaneously in the visible and near-infrared 
\citep{dominguez03, martinez_gonzalez06, dominguez06, marian08} with 
ground-based telescopes, and from space-bourne telescopes in the visible 
\citep{orozco_hinode07, lites08, bellot_orozco12}. There is a general consensus 
that the strength of the magnetic field lies in the hG regime, yet the 
observation of the Hanle effect at low spatial resolution shows that the 
magnetic energy stored in the quiet Sun is significant for the global energetics 
of the Sun \citep{trujillo_nature04}.

But the consensus is lost when one deals with the topology of the field. Some 
researchers conclude that the field has to be close to isotropic 
\citep{marian_clv08,bommier09,asensio_hinode09}, others conclude that the field 
is preferentially horizontal \citep{orozco_hinode07,lites08} and others show 
that stronger fields are preferentially vertical, becoming nearly isotropic in 
the weak flux density limit \citep{stenflo10}. The main reason for these 
apparent controversial results is that, at our best present observational 
capabilities and polarimetric sensitivity, we do not resolve individual magnetic 
structures.  The spatial organization of the magnetic fields in the quiet Sun is
very complex; we only hint organized, intermittent loop structures \citep[e.g.,][]{marian_deadcalm12} but they represent a small fraction 
of the surface. Although not yet demonstrated, it is tempting to consider that the rest appears as a multiscale stochastic 
medium, probably made of magnetic loops with scales below our resolution capabilities. 
Although intrinsically random, it is important to remember that a stochastic medium can appear highly ordered at many
scales. This is the case, for instance, of a stochastic process in scales (differences of the
properties at different times and/or positions) rather than purely in time or spatial
position. In such a case, the stochastic process is described by a probability distribution
that relates the differences in size between objects at one scale and at a smaller scale \citep[e.g.,][]{vankampen92,frisch95}.

As pointed out by \cite{asensio_hinode09}, one of the fundamental problems for 
inferring the statistical properties of the magnetic field vector in the quiet 
Sun resides in the large uncertainties in the inferred parameters induced
by the presence of degeneracies.
The situation is worsened by the presence of noise \citep{borrero_kobel11}. 
\cite{marian_dipolo12} and \cite{borrero_kobel12} showed that the inversion of
Stokes profiles with noise in Stokes $Q$ and $U$ leads to an artificial overpopulation of very inclined fields.


It is then crucial to have good estimations of the uncertainties on the inferred magnetic field vectors.
This is usually not the case when using standard inversion codes, independent of the approximation used to obtain the Stokes parameters. Error bars in least-squares inversion codes that use the Levenberg-Marquardt algorithm \citep{auer_heasly_house77,skumanich_lites87,lites_skumanich90,keller90,sir92,socas_trujillo_ruiz00,frutiger00}
are not precise in cases with degeneracies \citep[for the quiet Sun, see][]{martinez_gonzalez06}. 
The reason is that the errors are obtained approximating the
$\chi^2$ hypersurface with a hyperparaboloid, whose curvature matrix is given by the Hessian evaluated at the
location of the minimum. Although the error bars can be somehow patched \citep[see][]{sanchez_almeida_misma97}, they
are not precise at all. Therefore, it is important to carry out a fully Bayesian inference in which error
bars are correctly predicted for the model parameters in terms of the noise level in the observed
Stokes parameters and taking into account all the degeneracies and ambiguities.

Given the necessity to carry out the fully Bayesian inversion \citep{asensio_martinez_rubino07}, it is then not trivial how to extract the
\emph{general} properties of the magnetic field observed in a field-of-view (FOV). As usual in Bayesian inference, the solution to the inference
problem is given in probabilistic terms as a posterior distribution over the model parameters. Potentially, there is
such a posterior distribution for every observed pixel in the FOV, which is pointing out the
uncertainties in the model parameters which are a consequence of both the noise in the Stokes parameters and the
inherent ones. One could, for instance, take the mean of the posterior for each pixel and then carry out the
histogram of these means to build the distribution of a parameter of interest. This is, in essence, what has been
done in the past with standard inversion codes. However, this neglects the important information related to
the presence of noise and/or degeneracies and will potentially lead to biased distributions of the parameters.
For instance, the mean of a skewed distribution is biased and heavily influenced by the tails. Something similar happens with a very
broad distribution with no clear peak.

The hierarchical approach that we follow in this paper is the Bayesian way of propagating the pixel-by-pixel 
uncertainties to the distribution of the physical parameters on the FOV \citep{gelmanHierarchical07}. As explicited later, this hierarchical
approach is the equivalent to the statistical characterization of a parameter to which we
do not have direct observational access, but has to be inferred from observations. The fundamental
difficulty with this hierarchical approach is that the posterior distribution becomes very 
high-dimensional and can lead to computational problems when sampling it using a standard Markov
Chain Monte Carlo (MCMC) method. We have approximated the marginal posterior for the hyperparameters
using importance sampling.

\section{Hierarchical modeling of the quiet Sun}
In this section we propose to do a Bayesian analysis of the magnetism of the quiet Sun, correctly taking into account
all the ambiguities of the model, both the inherent and those produced by the presence of noise. We detail in
the following the model used to explain the signals and the hierarchical structure, with a detailed
description of the priors.

\subsection{Generative model}
\label{sec:generativeModel}
Our observables are
the wavelength variation of the Stokes parameters across a given spectral line, which we assume
are obtained using the following generative model:
\begin{align}
\mathbf{S}_\mathrm{obs}(\lambda) &= \mathbf{S}_\mathrm{syn}(\lambda;\thetabold)+\epsilonbold(\lambda),
\end{align}
where $\mathbf{S}(\lambda) = [I(\lambda),Q(\lambda),U(\lambda),V(\lambda)]^T$. For simplicity, we assume 
that $\epsilonbold(\lambda)$ represents uncorrelated Gaussian random noise, characterized by a variance $\sigma_n^2$.
The synthetic Stokes profiles, $\mathbf{S}_\mathrm{syn}(\lambda;\thetabold)$ depend on a set of
parameters, $\thetabold$, that will be defined in the following.

Current state-of-the-art parametric models of the solar magnetism are not able to capture
the organization of the field in the apparently stochastic solar atmosphere.
Strictly speaking, we should expect the distribution of fields in each pixel to be the result
of the addition of many scales simultaneously, with all scales probabilistically coupled. 
Until we study the quiet Sun equipped with such a model, we can
only aspire to grasp some general properties of it. That is what we do here, proposing a very
simple two-component model for explaining the signals, something that it is surely
far from the reality in many locations in the quiet Sun. Despite its simplicity, 
the model has been proved to explain a large fraction of the average polarimetric signals in the quiet Sun.

At a given pixel, we consider that the magnetic field 
strength is sufficiently weak so that the Stokes parameters can be assumed to be in the weak-field 
regime \citep[][]{landi73}. In such a regime, the Zeeman splitting $\Delta \lambda_B$ has to be much
smaller than the Doppler broadening, $\Delta \lambda_D$ \citep[e.g.,][]{landi_landolfi04}:
\begin{equation}
B < \frac{4 \pi m c}{\bar g \lambda_0 e_o} \sqrt{\frac{2kT}{M}+v_\mathrm{mic}^2},
\end{equation}
where $m$ and $e_0$ are the electron mass and charge, respectively, $c$ is the speed of light,
$k$ is the Boltzmann constant, $M$ is the mass of the species, $\lambda_0$ is
the central wavelength of the spectral line under consideration, $\bar g$ is the effective Land\'e factor and $v_\mathrm{mic}$ is the microturbulent
velocity. For the doublet of iron lines at $\lambda_0=630$ nm, using $v_\mathrm{mic}=1$ km s$^{-1}$
and $T=5800$ K, we end up with:
\begin{equation}
\bar g B < 1900 \mathrm{G}.
\end{equation}
Given that $\bar g$ ranges between 2.5 and 3 in the doublet lines at 630 nm, the weak-field can be applied
to field strengths up to $\sim 600-800$ G, although a complete calculation of the Stokes parameters show that
the weak-field approximation holds up to $\sim 1.2 kG$.

\begin{figure}
\centering
\includegraphics[width=0.85\columnwidth]{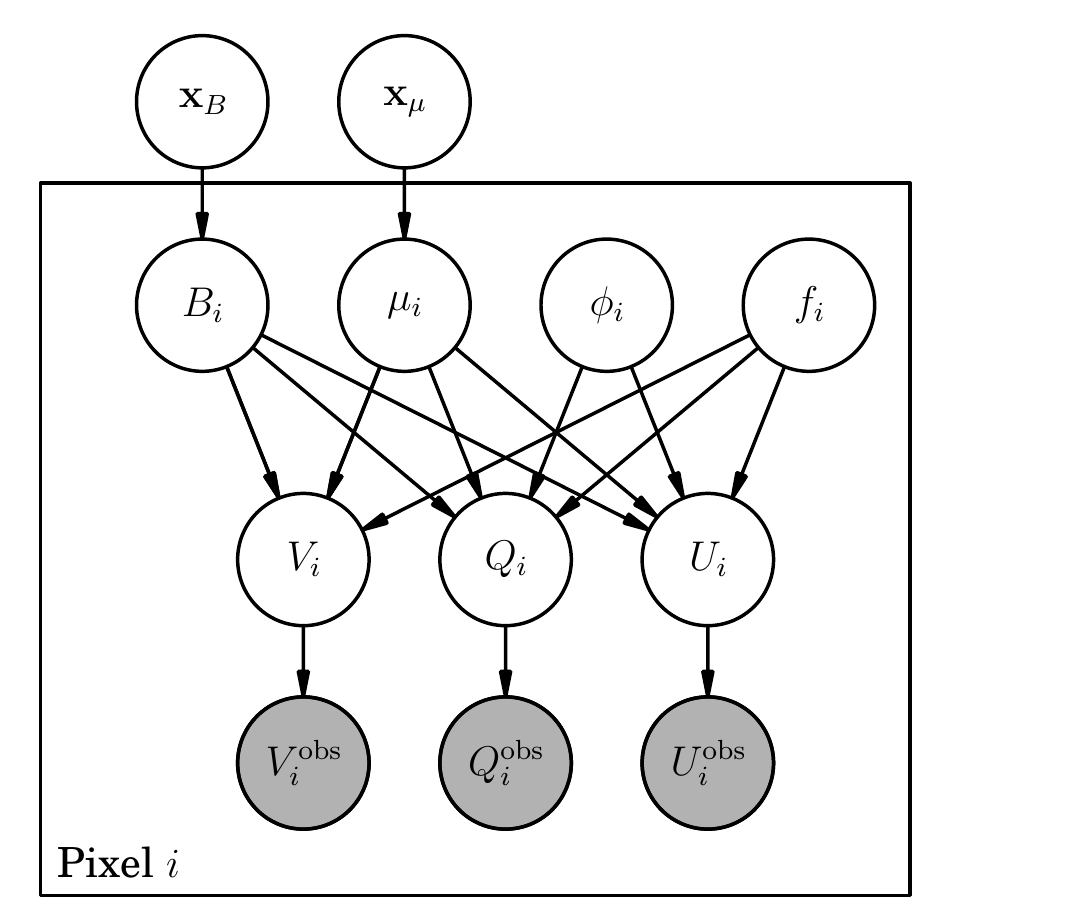}
\caption{Graphical model representing the hierarchical Bayesian scheme that we
used to analyze the set of Stokes signals in the quiet Sun. Open circles represent random
variables (note that both model parameters and observations are considered as
random variables), while the grey circle represents a measured quantity. The frame
labeled ``Pixel $i$'' represents that everything inside the frame has to be repeated for all
the observations. An arrow between two nodes illustrates dependency. 
The nodes that are outside the frame are the hyperparameters of the
model and are common to all pixels.}
\label{fig:graphical_model}
\end{figure}

Assuming that the weak-field approximation holds, the Stokes profiles can be explained with
the set of parameters $\thetabold=(B,\mu,f,\phi)$, where $B \in[0,\infty)$ is the magnetic field strength,
$\mu \in[-1,1]$ refers to the cosine of the inclination angle of the magnetic field vector with respect to the line-of-sight, and $\phi \in [0,2\pi]$ is the
azimuth of the magnetic field vector. Finally, $f \in [0,1]$ is the fraction of the resolution element that is filled
with magnetic field, with the remaining $1-f$ fraction being field-free. 
Following \citep{landi92}, we simplify the model by assuming that the presence of a magnetic field does not affect Stokes $I$, so that it
is the same in the magnetic and the non-magnetic fraction of the resolution element. 
All in all, the series expansion at first order in Stokes $V$ and second order in Stokes $Q$ and $U$
yields the following expressions for an arbitrary pixel $i$:
\begin{align}
V_i(\lambda) &= \alpha B_i \mu_i f_i \left( \frac{d{I}_i(\lambda)}{d\lambda} \right), \nonumber \\
Q_i(\lambda) &= \beta B_i^2 (1-\mu_i^2) f_i \cos 2\phi_i \left( \frac{d^2{I}_i(\lambda)}{d\lambda^2} \right), \nonumber \\
U_i(\lambda) &= \beta B_i^2 (1-\mu_i^2) f_i \sin 2\phi_i \left( \frac{d^2{I}_i(\lambda)}{d\lambda^2} \right),
\label{eq:synthetic_model}
\end{align}
where $\alpha=-4.67 \times 10^{-13} \lambda_0^2 \bar g$ and $\beta=-5.45 \times 10^{-26} \lambda_0^4 \bar G$
are constant that depend on the central wavelength of the spectral line and on the effective Land\'e factor, $\bar g$,
and its equivalent for linear polarization, $\bar G$. Additionally, we have particularized the model parameters to
the pixel of interest. As a requisite for the previous expressions
to hold, we have to additionally assume that the magnetic field vector is constant along the line of sight in the formation region of the spectral line.
A point that deserves a comment is the fact that the thermodynamic properties of the fraction $f$ of the pixel that generates
the Stokes $Q$, $U$ and $V$ signals are usually not the same of the remaining $1-f$ fraction. In such a case, the simple
approach followed in this paper cannot be applied and one has to be aware that there is some remaining information about 
the filling factor in the Stokes $I$ profile \citep[e.g.,][]{orozco_pasj07}. However, a careful analysis of this
more complicated case has to be carried out to avoid biasing the inferred values.

\subsection{Hierarchical probabilistic model}
\label{sec:hierarchical}
The statistical model that we use in this paper to extract information from the observations is displayed in
graphical form in Fig. \ref{fig:graphical_model}. All the variables inside open circles are considered
random variables\footnote{Random variables are used in Bayesian probability to model all sources of uncertainty. Therefore, any variable of the problem
that we do not know with infinite precision is considered to be a random variable: observations, model parameters, hyperparameters, etc.}, while those inside shaded circles are observations. The Stokes $Q$, $U$ and $V$ profiles of
pixel $i$ of the $N$ available are explained using the set of variables $\{B_i,\mu_i,\phi_f,f_i\}$ and the synthetic model of
Eq. (\ref{eq:synthetic_model}). For each pixel, the synthetic Stokes profiles are compared with the 
observed Stokes profiles using the appropriate model for the noise discussed in Sec. \ref{sec:likelihood}.
The hierarchical character of the model comes from the fact that we make the prior distribution of the parameters of the model $\{B_i,\mu_i\}$ with $i=1 \ldots N$
depend on a set of parameters $\mathbf{x}_B$ and $\mathbf{x}_\mu$ that, lying outside the frame, are 
thus common to all pixels\footnote{Note that
models with a large or even infinite number of parameters are routinely used to explain observations \citep[see, e.g.,][]{asensiomanso12}.}.
This is one of the improvements brought by the hierarchical approach, in contrast to the previous work in \cite{asensio_hinode09}, where the priors were 
chosen to be uninformative.

We do not consider hierarchy in the azimuth and the filling. Concerning the azimuth, we do it because
a simple statistical analysis of the $Q(\lambda)$ and $U(\lambda)$ signals yields that the azimuths are random in the quiet
Sun. For this reason, we will not get much information from this parameters. This affirmation is based on several
observations. For instance, a principal component analysis of the linear polarization signals in a large region of the quiet Sun
shows exactly the same first few principal components in both $Q(\lambda)$ and $U(\lambda)$. As a consequence, the prior
distribution is $p(\phi_i) = (\phi_\mathrm{max}-\phi_\mathrm{min})^{-1}$, where $\phi_\mathrm{min}$ and $\phi_\mathrm{max}$
are the limits of the parameter.
Concerning the filling factor, it is important to note that the application of the weak-field approximation to explain 
polarimetric signals coming from non-resolved structures is problematic because it is heavily degenerate with the other parameters.
Consequently, we will not be able to extract relevant information from it. Instead of just making $f=1$ and assuming
that the whole pixel is filled with a magnetic field, we consider it as a nuisance parameter that is integrated
during the marginalization. Therefore, a priori $p(f_i) = 1$.

Following the standard Bayesian formulation of an inference problem, the solution has to be given
in terms of the posterior distribution for all the parameters, that encodes all the information
about the parameters of interest. We represent it as 
$p(\mathbf{B},\mubold,\phibold,\mathbf{f},\mathbf{x}_B,\mathbf{x}_\mu|\mathbf{D})$,
where $\mathbf{D}=\{D_1,D_2,\ldots,D_N\}$ and $D_i=\{I_\mathrm{obs}^i(\lambda),Q_\mathrm{obs}^i(\lambda),U_\mathrm{obs}^i(\lambda),V_\mathrm{obs}^i(\lambda)\}$ refers to the 
set of Stokes profiles for $i$-th pixel of the set of $N$ observed pixels.
Likewise, $\mathbf{B}$, $\mubold$, $\phibold$ and $\mathbf{f}$ are vectors that contain all the parameters of the model 
for all the pixels, while $\mathbf{x}_B$ and $\mathbf{x}_\mu$ are the set of hyperparameters
that are used to describe the priors for the physical parameters of interest (see Sect. \ref{sec:priors}).
Applying the Bayes theorem, the posterior can be written as:
\begin{align}
p(\mathbf{B},\mubold,\phibold,\mathbf{f},\mathbf{x}_B,\mathbf{x}_\mu|\mathbf{D}) = \frac{
p(\mathbf{D}|\mathbf{B},\mubold,\phibold,\mathbf{f},\mathbf{x}_B,\mathbf{x}_\mu) 
p(\mathbf{B},\mubold,\phibold,\mathbf{f},\mathbf{x}_B,\mathbf{x}_\mu)}{p(\mathbf{D})},
\label{eq:bayes_theorem}
\end{align}
where $p(\mathbf{D})$ is the evidence, a normalization constant that is unimportant in our analysis.
In the previous equation, $\mathcal{L}=p(\mathbf{D}|\mathbf{B},\mubold,\phibold,\mathbf{f},\mathbf{x}_B,\mathbf{x}_\mu)$ is the likelihood, which measures the ability of
a set of parameters to fit the observations and that we will discussed in Sec. \ref{sec:likelihood}. Likewise, 
$p(\mathbf{B},\mubold,\phibold,\mathbf{f},\mathbf{x}_B,\mathbf{x}_\mu)$ is the prior distribution, that we elaborate in Sec. \ref{sec:priors}.

\subsection{Likelihood}
\label{sec:likelihood}
The likelihood of Eq. (\ref{eq:bayes_theorem}) can be simplified in two steps. First, it is clear from Eq. (\ref{eq:synthetic_model})
and Fig. \ref{fig:graphical_model}
that the synthetic Stokes profiles (and, consequently, the
likelihood) do only depend on the set of variables $\{\mathbf{B},\mubold,\phibold,\mathbf{f}\}$, and not on the hyperparameters $\mathbf{x}_B,\mathbf{x}_\mu$. Therefore,
the simplification $p(\mathbf{D}|\mathbf{B},\mubold,\phibold,\mathbf{f},\mathbf{x}_B,\mathbf{x}_\mu) = p(\mathbf{D}|\mathbf{B},\mubold,\phibold,\mathbf{f})$ applies. Second, we make
the assumption that the measurements for all the pixels are statistically independent, so that the likelihood factorizes as:
\begin{equation}
\mathcal{L} = \prod_{i=1}^N \mathcal{L}_i = \prod_{i=1}^N p(D_i|B_i,\mu_i,\phi_i,f_i).
\end{equation}

The analytical expression for each individual likelihood depends on the noise statistics.
If the observations are perturbed with Gaussian noise of variance $\sigma_n^2$ 
(we assume for simplicity that there is no correlation between different wavelengths or
different Stokes profiles, although it can be easily
generalized to the case in which such correlation matrix is known), the likelihood for a
single pixel is described by a Gaussian with zero mean and variance $\sigma_n^2$. Making everything explicit, 
each likelihood can be written as
\begin{align}
\mathcal{L}_i&=p(D_i|B_i,\mu_i,\phi_i,f_i) = (2\pi)^{-M/2} {\sigma_n}^{-M} \nonumber \\
&\times \exp \Bigg\{ -\frac{1}{2{\sigma_n}^2} \sum_{j=1}^M  \left[ V_i(\lambda_j) - \alpha_j B_i \mu_i f_i
\left( \frac{d{I}_i(\lambda)}{d\lambda} \right)_j \right]^2   \nonumber  \\
&- \frac{1}{2{\sigma_n}^2} \sum_{j=1}^M  \left[ Q_i(\lambda_j) - \beta_j B_i^2 (1-\mu_i^2) f_i \cos 2\phi_i
\left( \frac{d^2{I}_i(\lambda)}{d\lambda^2} \right)_j \right]^2 \nonumber  \\
&- \frac{1}{2{\sigma_n}^2} \sum_{j=1}^M  \left[ U_i(\lambda_j) - \beta_j B_i^2 (1-\mu_i^2) f_i \sin 2\phi_i
\left( \frac{d^2{I}_i(\lambda)}{d\lambda^2} \right)_j \right]^2
\Bigg\}
\label{eq:likelihood}
\end{align}
where $M$ is the number of wavelength points of each observed Stokes profiles. Further simplications in 
the notation are shown in Appendix \ref{sec:appendixA}.

\subsection{Priors}
\label{sec:priors}
The prior distribution encodes all the a-priori information that we know about the parameters. Instead of using
fixed prior distributions, in a hierarchical approach we make them depend on additional parameters (termed hyperparameters),
that are inserted in the Bayesian inference. The dependencies can be easily extracted from
the graphical model of Fig. \ref{fig:graphical_model}, so that the full prior distribution can be
factorized according to:
\begin{align}
p(\mathbf{B},\mubold,\phibold,\mathbf{f},\mathbf{x}_B,\mathbf{x}_\mu) &= p(\mathbf{B}|\mathbf{x}_B) p(\mubold|\mathbf{x}_\mu) p(\mathbf{x}_B) p(\mathbf{x}_\mu) p(\mathbf{f}) p(\phibold)
\end{align}
which can be even further simplified by assuming that the prior for the parameters of each pixel are 
independent, so that
\begin{align}
p(\mathbf{B},\mubold,\phibold,\mathbf{f},\mathbf{x}_B,\mathbf{x}_\mu) &= p(\mathbf{x}_B) p(\mathbf{x}_\mu) \nonumber \\
&\times \prod_{i=1}^N p(B_i|\mathbf{x}_B) p(\mu_i|\mathbf{x}_\mu)p(\phi_i) p(f_i).
\end{align}
Note that the prior for the azimuth is left non-hierarchical and chosen to be uniform in the interval $[0,\pi]$, while
that of the filling factor is also non-hierarchical and uniform in the interval $[0,1]$. Given the
inherent 180$^\circ$ ambiguity in the azimuth in the line-of-sight of the Zeeman effect, we decided to limit the
solution to only one of the solutions. This is motivated by two reasons. First, once the solution is obtained,
we immediately know the ambiguous solution. Second, working with multimodal distributions is
problematic and dealing with the two ambiguous solutions offers nothing new to the hierarchical analysis.


Using previous experience \citep{asensio_arregui13,asensio14}, we have decided to use very simple prior distributions for the
model parameters, chosen based on the conditions: i) they are mathematically simple but flexible
enough to adapt during the inference and, ii) they naturally fulfill the physical constraints. Note that the assumed prior is also an 
inherent part of the model, at the same level as the generative model.
The fact that the hyperparameters are random variables, will allow us to use these simple
prior distributions to generate quite complex global distributions. 

Given that $B \in [0,\infty)$, it
makes sense to use a log-normal prior for this lower-bounded parameter:
\begin{equation}
p(B_i|\mathbf{x}_B) = \mathrm{LN}(B_i;\alpha_B,\beta_B)= \frac{1}{\sqrt{2\pi} \beta_B B_i} \exp \left[-\frac{(\log B_i-\alpha_B)^2}{2\beta_B^2} \right],
\label{eq:prior_B}
\end{equation}
where $\alpha_B \in (-\infty,\infty)$ and $\beta_B>0$ are the hyperparameters. In the
notation used in Fig. \ref{fig:graphical_model}, we have that $\mathbf{x}_B=(\alpha_B,\beta_B)$.
One of the main properties of this prior is that, independently of the value of $\alpha$ and $\beta$, the probability of
having $B=0$ is zero. \cite{dominguez06b} and \cite{sanchezalmeida07} pointed out that, when the field strength is weak, the field 
becomes very tangled and random. Consequently, it is very improbable that the three components of the magnetic field vector become zero 
simultaneously, something that is naturally fulfilled by the prior.

Concerning the cosine of the heliocentric angle, is is limited to the bounded intervals $\mu \in [-1,1]$. 
A natural distribution for such bounded parameter which is able to take a large
variety of shapes is the scaled Beta prior:
\begin{align}
p(\mu_i|\mathbf{x}_\mu) &= \mathrm{Beta}(\mu_i;\alpha_\mu,\beta_\mu,a,b) \nonumber \\
&= \frac{(b-a)^{1-\alpha_\mu-\beta_\mu}}{B(\alpha_\mu,\beta_\mu)} (\mu_i-a)^{\alpha_\mu-1} (b-\mu_i)^{\beta_\mu-1},
\label{eq:prior_mu}
\end{align}
with $\alpha_\mu >0$ and $\beta_\mu>0$ the hyperparameters, $B(\alpha_\mu,\beta_\mu)=\Gamma(\alpha_\mu)\Gamma(\beta_\mu)/\Gamma(\alpha_\mu+\beta_\mu)$ the Beta function
\citep{abramowitz72} and $a=-1$ and $b=1$ are the limits of the interval. The hyperparameters will then be
$\mathbf{x}_\mu=(\alpha_\mu,\beta_\mu)$.

\subsubsection{Priors for hyperparameters}
Given that we have introduced four hyperparameters in the Bayesian inference, we have to use priors for them.
Concerning the prior for the magnetic field strength, we use the standard approach and set a Jeffreys' prior 
for the scale parameter $\beta_B$, while setting an uniform prior for the location parameter $\alpha_B$. For the Beta prior
for $\mu$, leaving uniform priors
for the hyperparameters of a Beta prior will surely lead to an improper posterior (the integral of the posterior
becomes infinity). \cite{gelman_bayesian03} suggest to use flat priors on the variables
\begin{equation}
\bar \mu=\frac{ \alpha_\mu b + \beta_\mu a}{\alpha_\mu+\beta_\mu}, \qquad \nu=(\alpha_\mu+\beta_\mu)^{-1/2},
\end{equation}
which are the mean of the distribution and inverse square root of the sample size. Using the Jacobian of the transformation from $(\bar \mu,\nu)$ to $(\alpha_\mu,\beta_\mu)$,
the prior becomes $p(\alpha_\mu,\beta_\mu)=(\alpha_\mu+\beta_\mu)^{-5/2}$.

\subsubsection{Marginal posterior}
From the previous considerations, we can obtain the full posterior by multiplying the likelihood and
the priors, yielding:
\begin{align}
p(\mathbf{B},&\mubold,\phibold,\mathbf{f},\mathbf{x}_B,\mathbf{x}_\mu|\mathbf{D}) \propto p(\mathbf{x}_B) p(\mathbf{x}_\mu) \nonumber \\
&\times \prod_{i=1}^N p(D_i|B_i,\mu_i,\phi_i,f_i) p(B_i|\mathbf{x}_B) p(\mu_i|\mathbf{x}_\mu) p(\phi_i) p(f_i).
\end{align}
The marginalization of all the individual parameters of the model for each pixel will yield
\begin{align}
p(\mathbf{x}_B,\mathbf{x}_\mu|\mathbf{D}) &\propto p(\mathbf{x}_B) p(\mathbf{x}_\mu) \nonumber \\
&\times \prod_{i=1}^N \Bigg[ \int \mathrm{d} B_i \, \mathrm{d}\mu_i \, \mathrm{d}\phi_i \, \mathrm{d}f_i p(D_i|B_i,\mu_i,\phi_i,f_i) \nonumber \\
&\times p(B_i|\mathbf{x}_B) p(\mu_i|\mathbf{x}_\mu) p(\phi_i) p(f_i) \Bigg].
\label{eq:final_posterior}
\end{align}
We finally note that either the integral over $\phi_i$ or over $f_i$ can be carried out analytically when
using flat priors. This reduces the dimensionality of the problem but is less general in case
one wants to use other priors. The expression for the marginal likelihood when integrating $f_i$ is shown in 
App. \ref{app:marginal_f}.

\begin{figure*}
\includegraphics[width=\textwidth]{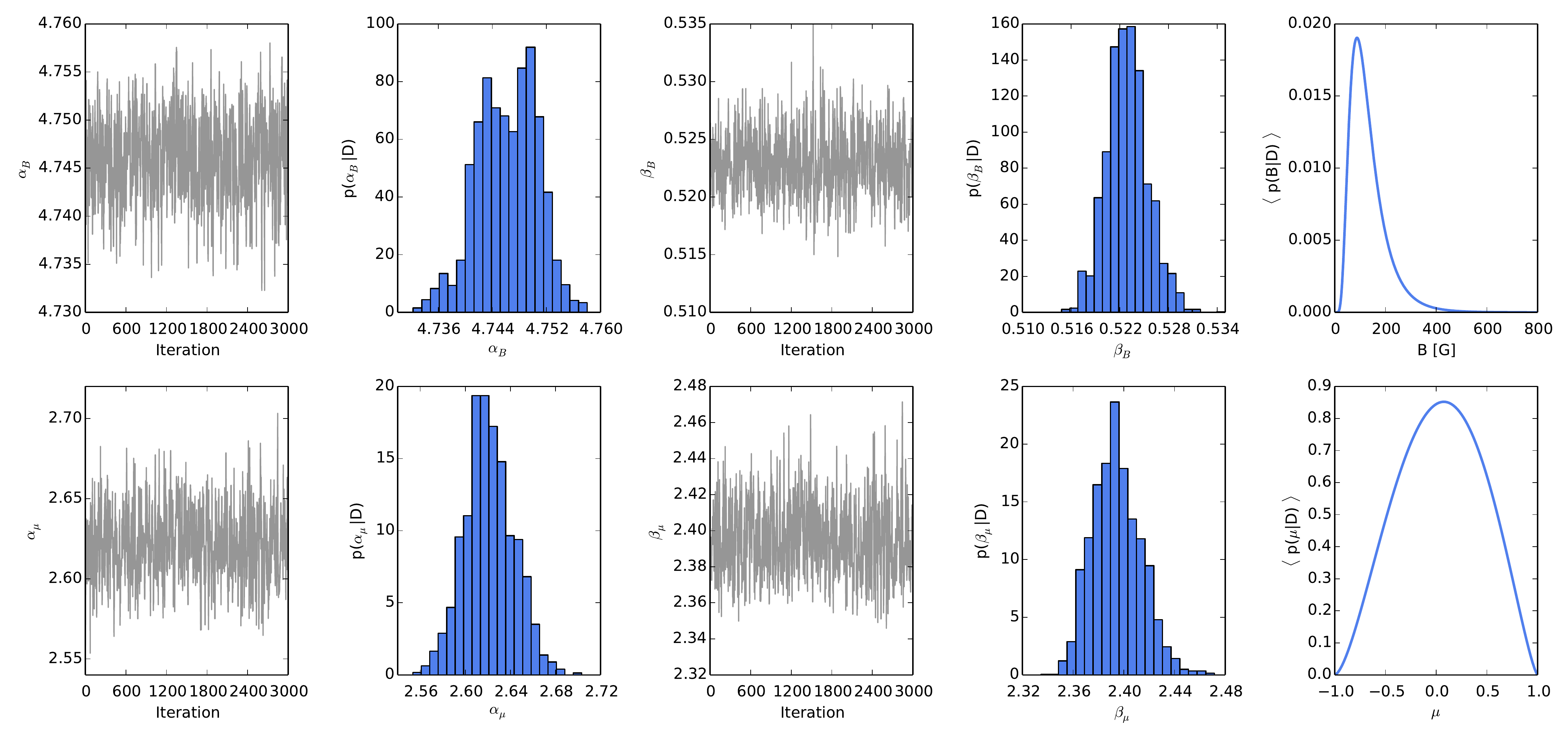}
\caption{Inferred values for the hyperparameters of the priors for $B$ (upper panels) and $\mu$ (lower panels). The
first and third columns show the last 3000 samples of the Markov chains, while the second and fourth show the
associated histograms. The last column displays the Monte Carlo inferred distribution of $B$ and $\mu$ taking into
account the observations. These results are obtained with 5\% of the FOV, although they remain the same 
as far as $\sim$0.5\% of the field of view is included in the analysis.}
\label{fig:hyperparameters}
\end{figure*}

\subsection{Inference}
Given the Stokes profiles observed at $N$ pixels, the posterior becomes a $4N+4$-dimensional distribution. It is usual
to apply MCMC methods to carry out the sampling from the posterior, but the large dimensionality
and the hierarchical character of the probabilistic model preclude an efficient solution even using
advanced techniques like Hamiltonian Monte Carlo methods \cite{hmc_duane87}. For this reason, we carry out an
efficient approximation for the marginal posterior of Eq. (\ref{eq:final_posterior}). The idea is based on carrying out the inference 
of each individual pixel independently using common priors $p(B_i)$ and $p(\mu_i)$, and then reconstructing back the results 
using importance sampling, similar to the approach used recently by \cite{hogg10} and \cite{brewer_hier14}:
\begin{align}
p(\mathbf{x}_B,\mathbf{x}_\mu|\mathbf{D}) &\propto p(\mathbf{x}_B) p(\mathbf{x}_\mu) \nonumber \\
&\times \prod_{i=1}^N \Bigg[ \int \mathrm{d} B_i \, \mathrm{d}\mu_i \, \mathrm{d}\phi_i \, \mathrm{d}f_i p(D_i|B_i,\mu_i,\phi_i,f_i) \nonumber \\
&\times \frac{p(B_i|\mathbf{x}_B) p(\mu_i|\mathbf{x}_\mu)}{p(\mathbf{B}) p(\mubold)} p(B_i) p(\mu_i) p(\phi_i) p(f_i) \Bigg].
\label{eq:final_posterior2}
\end{align}
If we carry out a sampling of the posterior for each individual pixel with the common priors, we can estimate the
integral using:
\begin{align}
p(\mathbf{x}_B,\mathbf{x}_\mu|\mathbf{D}) &\propto p(\mathbf{x}_B) p(\mathbf{x}_\mu) 
\prod_{i=1}^N \mathbb{E} \Bigg[ \frac{p(B_i|\mathbf{x}_B) p(\mu_i|\mathbf{x}_\mu)}{p(\mathbf{B}) p(\mubold)} \Bigg],
\label{eq:final_posterior_importance}
\end{align}
where $\mathbb{E}(x)$ refers to the expectation value, which is taken with respect to the pixel marginal posterior. Our calculations are done with flat common priors, with a 
large support for the prior for the magnetic field strength to ensure that it does not affect the computation of the hierarchical prior.

Summarizing, a standard MCMC\footnote{We use the Affine Invariant Markov chain Monte Carlo (MCMC) Ensemble sampler \texttt{emcee} developed by \cite{emcee12}.} sampling method is used to sample from the posterior of each
pixel and these samples are stored. For computational reasons, we only store 100 samples, which is enough to get a robust final result. Then, we sample 
from the marginal posterior of Eq. (\ref{eq:final_posterior_importance}) using again an MCMC. To this end, we compute,
at each iteration, the expectation inside the product with the stored samples \footnote{The code to reproduce the results in this paper
can be found in \texttt{https://github.com/aasensio/hierarchicalQuietSun}}.

\section{Results}
\label{sec:results}
The previous hierarchical model is applied to the quiet Sun observations presented by
\cite{lites08} using Hinode \citep{kosugi_hinode07} with the spectropolarimeter of the solar 
optical telescope\citep[SOT/SP;][]{lites_hinode01}. We focus on the Fe \textsc{i} line at $\lambda_0=6302.5$ \AA, which has
$\bar g=2.5$ and $\bar G=6.25$. Even though the dimensions of the map are enormous ($2048 \times 1024$ pixels, which cover an area
of $300'' \times 160''$ on the Sun), we have verified that 
the line-of-sight can be assumed to be roughly normal to the surface. Therefore, $\mu_i$ in our model will always represents the cosine of the 
angle that the magnetic field vector makes with the vertical.


We estimate the noise level in the continuum to be roughly the same for all the Stokes parameters and equal 
to $\sigma_n \sim 1.1 \times 10^{-3}$ in units of the continuum intensity \citep[e.g.,][]{lites08}. We only keep pixels 
that have signals in Stokes $Q$, $U$ or $V$ larger than 4.5 times the noise level in the continuum, so filtering
out pixels that only contain noise. After this filtering, only 27\% ($\sim$560000 pixels) of the FOV is considered.
When computing Eq. (\ref{eq:final_posterior_importance}), we test that the results are insensitive to the exact value of $N$ 
provided that $N \gtrsim 10000$.


We display in Fig.  the final results for $N=120000$ pixels.
The upper row shows the results for the hyperparameters of the magnetic field strength, while the
lower panels shows the final results for the hyperparameters of the cosine of the magnetic field inclination
with respect to the line-of-sight. The first and third columns show 3000 samples from the Markov chains
for the four hyperparameters, while the second and fourth columns display their histograms. They are the marginal posteriors for the
four hyperparameters. Note that they have the well-defined values 
$\alpha_B=4.75 \pm 0.01$, $\beta_B=0.52 \pm 0.01$, $\alpha_\mu=2.63 \pm 0.02$ and $\beta_\mu=2.40 \pm 0.02$. Using these values
and the properties of the log-normal distribution, we find $\langle B \rangle=132 \pm 1$ G and $\sqrt{\langle B^2 \rangle}=151 \pm 1$ G,
in agreement with previous results, which were obtained following different approaches \citep[e.g.,][]{trujillo_nature04,marianPhD06,dominguez06b}.

Using these results, there are two ways to compute
the global magnetic properties of the analyzed pixels, i.e., the distribution of field strengths and inclinations in
the whole FOV analyzed that would be compatible with noise and degeneracies in each pixel. The first one is to use what is known as the
type-II maximum likelihood approximation. To this end, we simply evaluate the priors
defined in \S\ref{sec:priors} at the most probable values of their parameters, obtained from the peaks on
Fig. \ref{fig:hyperparameters}. The second way, that is the one we use for producing the plots on the
last column of Fig. \ref{fig:hyperparameters}, is to compute the following Monte Carlo estimation of
the marginalization of the hyperparameters from the priors using $N_s$ samples:
\begin{align}
\langle p(B|D) \rangle &= \frac{1}{N_s} \sum_{i=1}^{N_s} \mathrm{LN}(B;\alpha_B^i,\beta_B^i)  \nonumber \\
\langle p(\mu|D) \rangle &= \frac{1}{N_s} \sum_{i=1}^{N_s} \mathrm{Beta}(\mu;{\alpha_\mu}^i,{\beta_\mu}^i,-1,1).
\end{align}
Given that the marginal posterior distributions for the hyperparameters are very well defined, the
type-II maximum likelihood and the Monte Carlo estimation essentially overlap.
The distributions in the rightmost panels of Fig. \ref{fig:hyperparameters}
are calculated taking into account the information from $N=120000$ pixels and their uncertainties
in the inferred parameters and combining them in one distribution. This includes the effect of
noise, degeneracies and any other uncertainty.
They constitute the main result of this paper and have to be confronted with previous studies.
Because of the presence of noise and degeneracies, these distributions broaden with respect to the
true ones. Observations with a better
signal-to-noise ratio would reduce this broadening but would never reduce the broadening produced by the inherent degeneracies of
the model.

\begin{figure}
\includegraphics[width=\columnwidth]{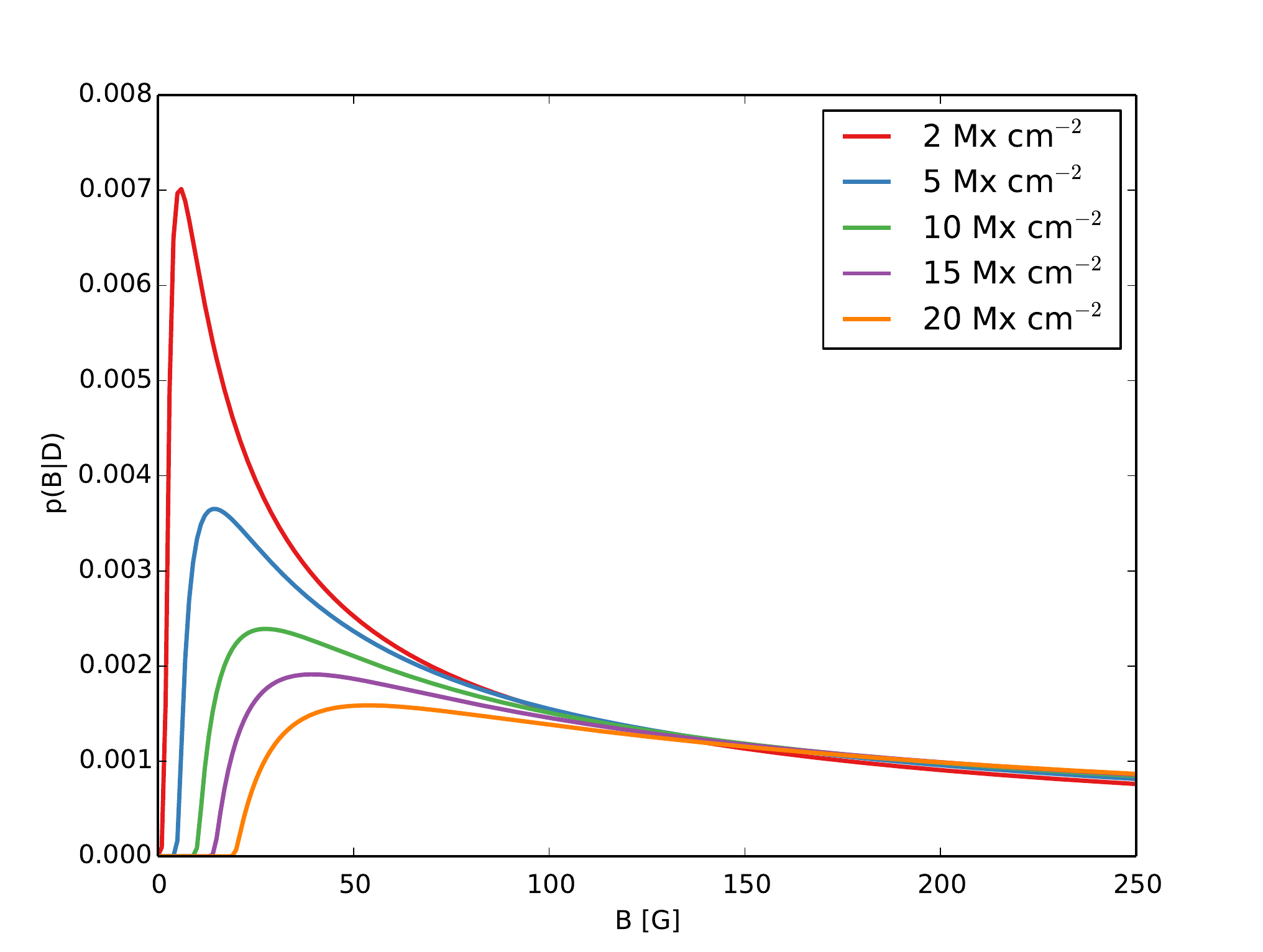}
\caption{Marginal posterior for the magnetic field strength when we measure different values of the product $fB$ (indicated
in the legend) with an uncertainty of 0.5 Mx cm$^{-2}$. The fact that we assume a flat prior for $f$ induces that
the most probable value for $B$ is very small, close to what one would measure assuming $f=1$. The extended tail
is produced by the small possible values of $f$.}
\label{fig:marginal_example}
\end{figure}

The distribution for the magnetic field strength is quite similar to previous results. 
This means that the effect of noise and degeneracies in earlier works
with histogramming the maximum-likelihood estimations was small. The
reason for that is that the amount of hG fields is so large that the influence
of the tails induced by the presence of noise and degeneracies is not very
important. Therefore, just picking up the mode of the distribution results in a
good estimation of the global distribution of magnetic field strengths.
The good point of the Bayesian approach is that we verify that taking the mode seems to be
a good strategy.
Fields are in the hG regime, with the distribution peaking around 85 G. 
With the observed data and our current analysis, the field strength is below 275 G with 95\% credibility.
Certainly, this should not be confused with the fraction of pixels having these fields, which would be a frequentist
Because we are including the filling factor and the inclination of the field into the inference, we are able to extract
values of the magnetic field strength, even though we are working in the weak-field regime.
In fact, we are considering all possible values of the filling factor and inclination that are
compatible with the observations for each individual pixel. Even in the extreme case that only the 
amplitude of circular polarization is available (a map of magnetic flux density), it is still possible 
to give inferences about the magnetic field strength. If we observe a certain magnetic flux density, $\Phi_\mathrm{obs}$,
and we model it with the simple expression $\Phi=Bf\mu$, the
marginal posterior for the field strength is given by
\begin{align}
p(B|\Phi) \propto p(B) \int_0^1 \mathrm{d}f \int_{-1}^1 \mathrm{d}\mu \exp \left[-\frac{\left( \Phi_\mathrm{obs}-Bf\mu \right)^2}{2\sigma_n^2} \right],
\end{align}
where we have assumed flat priors for $\mu$ in the interval $[-1,1]$ and $f$ in the interval $[0,1]$.
For a noise with standard deviation $\sigma_n=0.5$ Mx cm$^{-2}$, the marginal posteriors for a few measured values of $\Phi$ are displayed in 
Fig. \ref{fig:marginal_example}. According to this result, since the prior distributions for $f$ and $\mu$ are
assumed to be flat, the peak of the distribution or marginal maximum a-posteriori (MMAP) value of the magnetic field is small (close to $\Phi$), 
with a very long tail towards higher values. 

\begin{figure}
\includegraphics[width=\columnwidth]{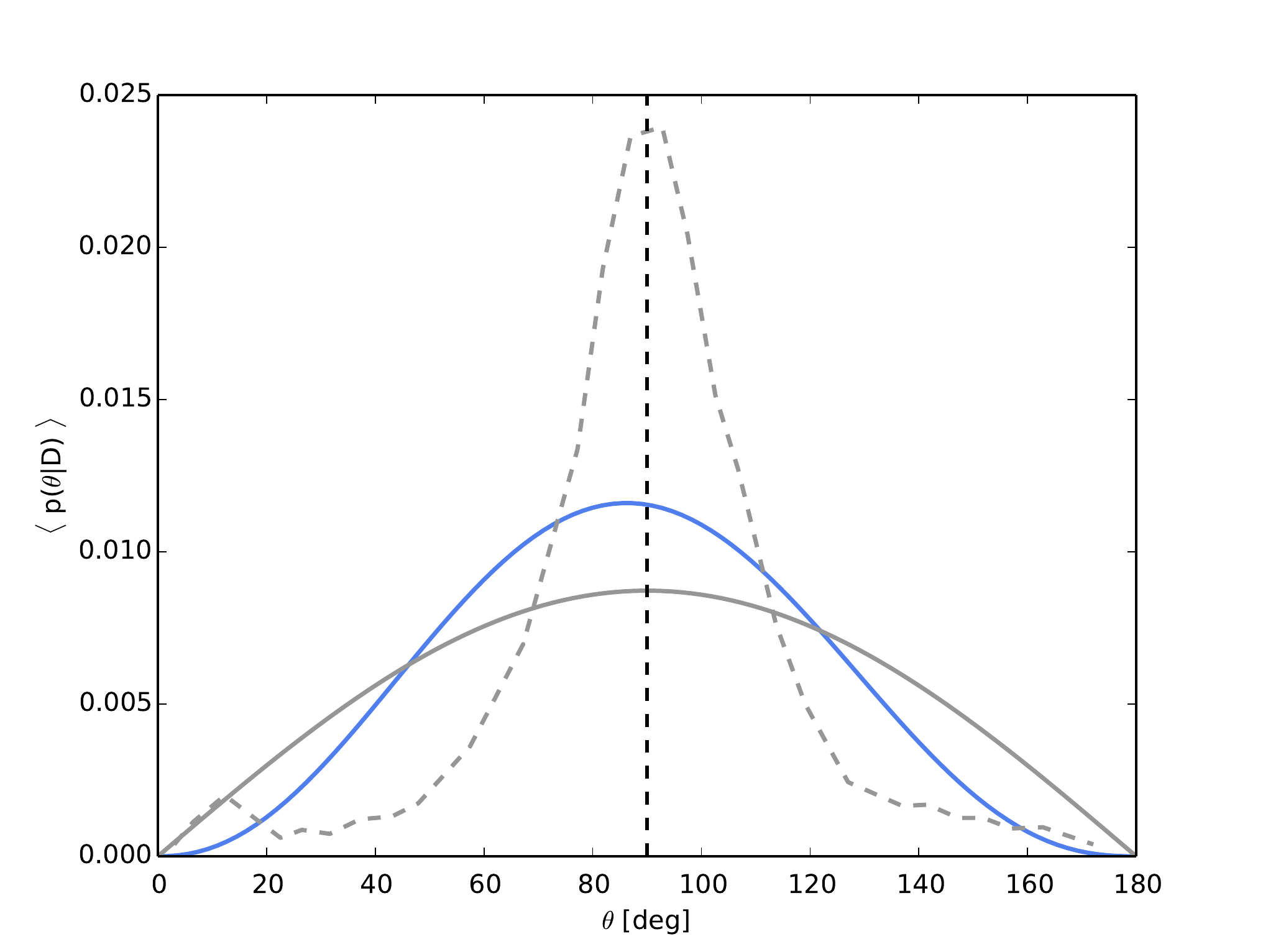}
\caption{Inferred global distribution of inclinations from the data (blue), compared with the expected
distribution of inclinations for an isotropic vector field (solid gray). For comparison, we have overplotted
in dashed gray line the distribution inferred by \cite{bellot_orozco12}, which we obtained by scanning
the original figure. For reference, the vertical dashed line indicates purely horizontal fields.}
\label{fig:inclination}
\end{figure}


Concerning the magnetic field inclination, a simple change of variables can be used to transform the 
distribution for $\mu$ shown in the lower panel of the last column of Fig. \ref{fig:hyperparameters} 
into a distribution for the inclination of the field, $\theta$. Starting from $p_\mu(\mu)$, the distribution 
$p_\theta(\theta)$ in terms of $\theta$ is given by
\begin{equation}
p_\theta(\theta) = p_\mu(\mu=\cos \theta) |\sin \theta|,
\end{equation}
where we have made explicit that the distributions are different. Applying this change of variables, we 
end up with the distribution displayed in blue in Fig. \ref{fig:inclination}. For comparison, we show the 
distribution associated with an isotropic field, which has $p_\mu(\mu)=1/2$ or, equivalently, $p_\theta(\theta)=\sin (\theta)/2$.
Additionally, we also display the distribution of inclinations obtained by \cite{bellot_orozco12} in dashed gray lines.
The distribution of fields is close to $\langle p(\theta|D) \rangle \propto \sin^2 \theta$, although with a slight skewness towards fields
pointing downwards ($\mu<0$, equivalently, $0^\circ<\theta<90^\circ$). Figure \ref{fig:inclination} shows that there are slightly more 
highly-inclined fields and slightly less highly-vertical fields than what one would expect for an isotropic field. In numbers,
the inferred distribution contains $\sim$80\% of the fields in the very inclined regime ($45^\circ<\theta<135^\circ$), while the 
isotropic distribution contains $1/\sqrt{2} \sim 70$\% in this regime. 

Our new result is to be preferred over our previous result in \cite{asensio_hinode09}, although with 
some caveats because we are using a simpler model for explaining the polarimetric signals. 
In \cite{asensio_hinode09}, the peak in 40$^\circ$ and 140$^\circ$ is a consequence of the
fact that the polarimetric signal is very weak, so only the polarity can be estimated in
many pixels. As a consequence, the median value that we used to summarize the marginal posteriors peak around the center of the 
intervals $[0^\circ,90^\circ]$ and $[90^\circ,180^\circ]$, as already explained in \cite{asensio_hinode09}. Additionally, 
given the scarcity of information, they also used the cumulative distribution to give
a hint that the field seems to be close to isotropic for the the pixels with the weakest signals.
In this updated work, we take these large uncertainties into account and extract
the global distribution of inclinations, under the assumption that all the pixels share a common
probability distribution. Note that, even though fields close to 90$^\circ$ are favored, the distribution is very close
to isotropic.
Finally, as compared with the distribution inferred previously by \cite{bellot_orozco12} from the same data, it is obvious that our results point towards a much
more quasi-isotropic distribution, in a way similar to the results of \cite{asensio_hinode09} and \cite{stenflo10}.

Even though the presence of noise complicates the inference of the field inclination, our results are certainly less affected than
other previous results for one reason: we compute all field inclinations that
are compatible with the observations for each individual pixel, together with
their associated probability. Then, these distributions are used to estimate
the global field inclination distribution, fully taking into account the
presence of uncertainties. If the noise variance is decreased in future observations, the ensuing posterior
distributions for each individual pixel will certainly be narrower, resulting in more informative
global field inclination distributions.

\section{Conclusions}
This paper presents our first attempt to infer global distributions of magnetic field strength and inclinations
from spectropolarimetric data taking fully into account all the degeneracies. To this end, we applied
a Bayesian hierarchical model. The difficulty of the statistical model forced us to use the weak-field
approximation simplified model to explain the polarimetric signals. Although simplified, this model
captures a large fraction of the behavior that is explained by more complicated models.

Our results indicate that the magnetic field strength has to be weak, below 275 G with 95\% credibility. This is
a direct consequence of the fact that we consider that all values of the filling factor in the interval $[0,1]$ are
equiprobable a-priori. Concerning the distribution of field inclinations, we find a rather quasi-isotropic
distribution, roughly proportional to $\sin^2 \theta$.

In the future, we plan to extend our hierarchical approach to more complicated models for the Stokes profiles.
The main obstacle resides on the potentially high dimensionality of the probabilistic model given that more
complicated models need a larger number of free parameters.

\begin{acknowledgements}
The diagram of Fig. \ref{fig:graphical_model} has been made with Daft (\texttt{http://daft-pgm.org}), developed by
D. Foreman-Mackey and D. W. Hogg.
Financial support by the Spanish Ministry of Economy and Competitiveness 
through projects AYA2010--18029 (Solar Magnetism and Astrophysical Spectropolarimetry) and Consolider-Ingenio 2010 CSD2009-00038 
are gratefully acknowledged. AAR also acknowledges financial support through the Ram\'on y Cajal fellowships.

\end{acknowledgements}


\onecolumn
\begin{appendix}
\section{Likelihood}
\label{sec:appendixA}
The likelihood of Eq. (\ref{eq:likelihood}) can be written in a more simplified form showing that
only 8 numbers per pixel are needed from the observations. Regrouping terms, the definition of the likelihood of Eq. (\ref{eq:likelihood}) 
can be simplified to read:
\begin{align}
p(D_i|B_i,\mu_i,\phi_i,f_i) &= (2\pi)^{-M/2} {\sigma_n}^{-M} 
\exp \Bigg\{ -\frac{1}{2{\sigma_n}^2} \left[ C_{V1i} + B_i^2 \mu_i^2 f_i^2 C_{V2i} - B_i \mu_i f_i C_{V3i} \right. \nonumber \\
&+ C_{Q1i} + B_i^4 (1-\mu_i^2)^2 f_i^2 \cos^2 2\phi_i C_{Q2i} - B_i^2 (1-\mu_i^2) f_i \cos 2\phi_i C_{Q3i} \nonumber \\
&+ \left. C_{U1i} + B_i^4 (1-\mu_i^2)^2 f_i^2 \sin^2 2\phi_i C_{U2i} - B_i^2 (1-\mu_i^2) f_i \sin 2\phi_i C_{U3i} \right]
\Bigg\},
\end{align}
where
\begin{align}
C_{V1i} =& \sum_{j=1}^M V_i^2(\lambda_j), \qquad C_{V2i} = \alpha^2 \sum_{j=1}^M \left( \frac{d{I}_i(\lambda)}{d\lambda} \right)^2_j, 
\qquad C_{V3i} = 2 \alpha \sum_{j=1}^M V_i(\lambda_j) \left( \frac{d{I}_i(\lambda)}{d\lambda} \right)_j \nonumber \\
C_{Q1i} =& \sum_{j=1}^M Q_i^2(\lambda_j), \qquad C_{Q2i} = \beta^2 \sum_{j=1}^M \left( \frac{d^2{I}_i(\lambda)}{d\lambda^2} \right)^2_j, 
\qquad C_{Q3i} = 2 \beta \sum_{j=1}^M Q_i(\lambda_j) \left( \frac{d^2{I}_i(\lambda)}{d\lambda^2} \right)_j \nonumber \\
C_{U1i} =& \sum_{j=1}^M U_i^2(\lambda_j), \qquad C_{U2i} = \beta^2 \sum_{j=1}^M \left( \frac{d^2{I}_i(\lambda)}{d\lambda^2} \right)^2_j, 
\qquad C_{U3i} = 2 \beta \sum_{j=1}^M U_i(\lambda_j) \left( \frac{d^2{I}_i(\lambda)}{d\lambda^2} \right)_j.
\label{eq:C_parameters}
\end{align}
Note that $C_{Q2i}=C_{U2i}$, which reduces the number of quantities needed to describe the information that we need
from the Stokes profiles to 8.

\section{Likelihood integrating the filling factor}
\label{app:marginal_f}
Given that the likelihood is factorizable and Gaussian, the filling factor can be marginalized from
the posterior analytically. This is possible if we use a flat prior for this parameter, so that:
\begin{align}
\int \mathrm{d}f_i p(D_i|B_i,\mu_i,\phi_i,f_i) p(f_i) = \frac{\sqrt{\pi}}{2\sqrt{C}} \exp \left[ - A+\frac{B^2}{4C} \right] 
\left[ \mathrm{erf} \left( \frac{B}{2\sqrt{C}} \right) - \mathrm{erf} \left( \frac{B-2C}{2\sqrt{C}} \right) \right],
\end{align}
where the quantities $A$, $B$ and $C$ are defined as:
\begin{align}
A &= C_{V1i}+C_{Q1i}+C_{U1i} \nonumber \\
B &= B_i \mu_i C_{V3i} + B_i^2 (1-\mu_i^2) \left( \cos 2\phi_i C_{Q3i} + \sin 2 \phi_i C_{U3i} \right) \nonumber \\
C &= B_i^2 \mu_i^2 C_{V2i} + B_i^4 (1-\mu_i^2)^2 \left( \cos^2 2\phi_i C_{Q2i} + \sin^2 2 \phi_i C_{U2i} \right)
\end{align}

\end{appendix}

\end{document}